\documentclass[12pt]{iopart}

\usepackage{amssymb,graphicx}

\begin{document}

\title[Electronic properties of corrugated graphene:  the Heisenberg principle]{Electronic properties of corrugated graphene, the Heisenberg principle and wormhole geometry in solid state}

\author{Victor Atanasov}
\address{Department of Condensed Matter Physics, Sofia University, 5 boul. J. Bourchier, 1164 Sofia, Bulgaria}
\ead{\mailto{vatanaso@gmail.com}}

\author{Avadh Saxena}
\address{Theoretical Division and Center for Nonlinear Studies,
Los Alamos National Laboratory, Los Alamos, NM 87545 USA}
\ead{\mailto{avadh@lanl.gov}}

\begin{abstract}
Adopting a purely two dimensional relativistic equation  for graphene's carriers contradicts the  Heisenberg uncertainty principle since it requires setting off-the-surface coordinate of a three-dimensional wavefunction to zero. Here we present a theoretical framework for describing graphene's massless relativistic carriers in accordance with this most fundamental of all quantum principles. A gradual confining procedure is used to restrict the dynamics onto a surface and normal to the surface parts and in the process the embedding of this surface into the three dimensional world is accounted for. As a result an invariant geometric potential arises in the surface part which scales linearly with the Mean curvature and shifts the Fermi energy of the material proportional to bending. Strain induced modification of the electronic properties or ``straintronics" is clearly an important field of study in graphene. This opens a venue to producing electronic devices, MEMS and NEMS where the electronic properties are controlled by geometric means and no additional alteration of graphene is necessary. The appearance of this geometric potential also provides us with clues as to how quantum dynamics looks like in the curved space-time of general relativity. In this context, we explore a two-dimensional cross-section of the wormhole geometry realized with graphene as a solid state thought experiment.
\end{abstract}

\pacs{71.10.Pm, 02.40.-k, 72.80.Vp}

\maketitle

\section{Introduction}

The two-dimensional allotrope of carbon, namely graphene, has emerged as a very promising electronic material.  In monolayer 
graphene electrons are described by the Dirac equation \cite{neto2}. Here we analyze the geometrically induced  chemical potential due to confinement (i.e. the gauge field) in the Dirac equation on a two dimensional sheet since  graphene is a two dimensional sheet.  Height fluctuations of a graphene layer (carriers are described by a relativistic equation) on a rough substrate generate a nonvanishing geometric potential. In this case, the shape of the graphene sheet is determined by a competition between the interaction of the layer with the rough substrate $H_{sub}$ (which tends to impose a preferred height), the elastic properties of the layer $H_{elast}$  and the electronic energy  $H_{el}$, which is also a function of geometry:
\[
H= H_{sub} + H_{elast} + H_{el} .
\]

The confining procedure employed here for exploring the electronic properties of graphene goes beyond the standard analysis of the Dirac equation in curved geometries \cite{vozmediano}. Here we will employ an alternative approach which is consistent with the Heisenberg principle. On one hand the starting point is the realization that graphene as a one-atom-thick membrane has carriers confined in a two dimensional space trapped in three dimensions. Furthermore the electrons are described by a massless relativistic equation. On the other hand the wavefunction of a quantum  particle is always three dimensional due to the Heisenberg uncertainty principle which forbids setting any of the coordinates to zero [this would lead to indefiniteness in the momentum according to $\Delta p \ge {\hbar}/{(2 \Delta x})$].  In this way any two dimensional quantum motion would have an evanescent off-surface component of the wavefunction which can probe the two dimensional surface for curvature \cite{9, daCosta}.  If the graphene sheet is curved, either intrinsically or extrinsically, then the carriers will at least be able to ``feel"  this curvature in the form of some mass or effective gauge field.

In fact, curvature of the sheets builds strain and from a microscopic point of view strain has been shown to  modify the electronic structure \cite{guinea}. This is a confirmation  of the line of thinking presented in this paper. When interatomic distances are modified due to strain in the underlying lattice caused by curvature, the periodicity is disrupted and the conditions for the application of Bloch theorem do not hold. Therefore, standard methods of solution fail to grasp the complexity of the rippling property of graphene.  Effectively this means that one has to use a constraining procedure which starts from three dimensions and gradually confines the quantum dynamics onto the surface. Note, for the Schr\"odinger equation the curvature induces an effective (da Costa) potential \cite{daCosta} through a confining procedure reducing the dimensionality of the quantum system. One question stands out: what is the corresponding potential for the relativistic case?  We have answered this question and illustrated it through graphene. In short, the effective potential has a linear dependence on the Mean curvature ($M$) as opposed to $M^2-K$, where $K$ is the Gaussian curvature, in the usual (or nonrelativistic) case. 

%%% new paragraph %%%%

This effect is studied by expanding around the two dimensional space of the graphene sheet for vanishingly small excursions in the third direction. Such a procedure is well known for the Schr\"odinger equation \cite{daCosta} and we have carried it out in the case of a relativistic equation. The idea behind this confinement is that an ``external force" gradually compresses the quantum dynamics onto a surface. In the following section we discuss where this force appears from in the case of graphene. When the excursion in the normal to the surface direction becomes small enough one can take the limit (the normal to the surface coordinate goes to zero) and split the original 3+1 dimensional relativistic equation into 2+1 and 1+1 dimensional equations. The 2+1 dimensional equation encodes the quantum dynamics onto the surface and the 1+1 dimensional equation describes the behavior of the ``evanescent component", that is the normal to the surface component of the wavefunction. In this way the Heisenberg principle is not violated since in the process described above no setting-to-zero of any coordinate of a three dimensional wavefunction takes place.

%%%%%%%%%%%%%%%

In reducing the Dirac equation from 3+1 dimensions to 2+1 and 1+1 dimensional equations we use the fundamental 2 representation of the Clifford algebra \cite{fulling}, which is appropriate for the isospin of graphene. The results of this confining procedure (Sec. II) and its specific applications (Sec. III) are also relevant for other fields of physics where phenomena are described within the framework of Dirac theory: chiral spin liquids, quantum Hall systems, anyons. 

Experimental confirmations, as to what is the correct relativistic description of quantum motion on a curved manifold, are important. Graphene presents with an experimental opportunity to answer this question with potential applications to the theory of gravity and its connection to quantum mechanics. This aspect also motivates us to consider the two-dimensional cross-sections of a wormhole geometry (Sec. IV).  Our main findings are summarized in Sec. V. 

\section{Confinement}
 
Now let us consider constrained nonrelativistic dynamics described by the Schr\"odinger equation with the extra chemical potential due to confinement\cite{daCosta}
\begin{equation}
V=-\frac{\hbar^2}{8m}(\kappa_1 - \kappa_2)^2= -\frac{\hbar^2}{2m} \left( M^2 -K \right) , 
\end{equation}
where $\kappa_i$ are the principal curvatures of the surface, $\hbar$ is the Planck's constant and $m$ is the effective mass. Here $M$ is the Mean curvature and $K$ is the Gauss curvaure. Note, by virtue of its derivation, this potential stems from the kinetic part of the constrained Schr\"odinger equation, that is in the case of the massless Dirac equation geometric potentials stemming from the kinetic part will not vanish.

As is well known, the Dirac equation is a ``square root"
 of the Schr\"odinger equation, meaning that whatever the constrained Dirac equation is, we would expect the chemical potential to scale linearly with the curvature since it scales as the square of the curvature in the case of the Schr\"odinger equation. 
 
 \begin{figure}[ht]
\begin{center}
\includegraphics[scale=0.4]{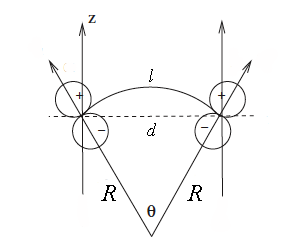}
\caption{\label{fig:overlap} Bending the surface of graphene by a radius $R$ and its
effect on the $\rm p_z$ orbitals. }
\end{center}
\end{figure}

Consider bending the graphene sheet by a radius $R.$ This will decrease
the distance between the orbitals from $l=R\theta$ to $d=2R \sin \left( \frac{\theta}{2} \right)$, 
\begin{equation}
l-d=l \left[ 1 -\frac{\sin x}{x}  \right]\approx l \frac{x^2}{6}, \qquad x\simeq \frac{l}{2R} \ll 1 . 
\end{equation}
The decrease in the distance between the orbitals increases the overlap
between the two lobes of the $\rm p_z$ orbital, see Fig. \ref{fig:overlap}. The overlap scales as the square of the curvature $1/R$ but this does not necessarily point to an effective chemical potential proportional to the square of the curvature  in the Dirac equation for the carriers\cite{neto1, neto2}.

Next, let us proceed with the derivation. Here we choose to use gauge covariant Dirac description for the massless carriers in graphene, both electrons and holes:
 \begin{equation}\label{eq:dirac3d}
 \left[ -i\hbar v_F \sum \gamma^a D_a + m(q_3)v_F^2 + V_{\perp}(q_3) \right] \Psi(t, q_i )=0 ,
\end{equation}
 where $(q_1, q_2)$ are the generalized coordinates of the surface and $q_3$ measures the deviation from the surface in the embedding 3 dimensional space. Here $D_a$ stands for a 3+1 dimensional gauge covariant derivative, $\gamma^a$ are curved gamma matrices $\gamma^a=\gamma^a(q_1, q_2, q_3)$ related to the usual constant matrices $\gamma^\mu$ by
 \begin{equation}
 \gamma^a(q_1, q_2, q_3)=e^a_\mu (q_1, q_2, q_3) \gamma^\mu,  
 \end{equation}
 where $e^a_\mu$ is the tetrad constructed from the metric tensor. 
 Here $ m(q_3)$ is an effective mass term with the following property
\begin{equation}
 m(q_3)=\left\{\begin{array}{ll}  0 & q_3=0  \\ m & q_3 \neq 0 \end{array}\right.
\end{equation}
and $V_{\perp}(q_3)$ is the constraining potential forcing the system onto the surface
\begin{equation}
V_{\perp}(q_3)=\left\{\begin{array}{ll}  0 , & q_3=0  \\ \infty & q_3 \neq 0 \end{array}\right\} . 
\end{equation}
Here a possible model of the confining potential is the charged plate potential
\begin{equation}
V_{\perp}(q_3) \sim |q_3|
\end{equation} 
since when the electron deviates from the two dimensional world of the sheet it disrupts the charge balance and a Coulomb force emerges.

The gauge covariant derivative is defined through its action on a vector field as follows
\begin{equation}\label{eq:D}
D_0=\partial_t, \qquad D_\sigma v^b = \partial_\sigma v^b + \Gamma_{\sigma j}^b v^j - i \frac{e}{c}A_\sigma v^b,
\end{equation}
where $A_{j}$ are the components of a vector potential and $\Gamma_{kl}^j$ are the Christoffel symbols of the  second kind
$
\Gamma_{kl}^j=\frac12 G^{jm} \left( G_{mk,l} + G_{ml,k} - G_{kl,m}  \right).
$ Here $G_{mk}$ is the metric tensor of the embedding space. This equation is defined in the entire embedding space $\vec{X}(q_1, q_2, q_3).$ 

Our goal now is to constrain (\ref{eq:dirac3d}) onto a surface $\vec{x}(q_1, q_2)$ and separate in two parts: one that describes the relativistic dynamics on the surface and one that describes the dynamics in a direction $q_3$ orthogonal to the surface. We will treat the off surface dynamics nonrelativistically since as soon as the carriers leave the two dimensional world of the graphene sheet they attain mass (an adjustable parameter in the confinement model) and the hexagonal symmetry of the underlying lattice no longer applies indicating the nonrelativistic limit. This also indicates that the off-surface component of the wavefunction is a scalar which allows separation of variables.

To achieve the above goal we will minimize the action of the Dirac equation (\ref{eq:dirac3d})
\begin{eqnarray}
\nonumber S&=&\int dV \; \overline{\Psi} \left[ -i\hbar v_F \sum \gamma^a D_a + m(q_3)v_F^2 + V_{\perp}(q_3) \right] \Psi,
\end{eqnarray}
where the conjugated two component spinor wavefunction is $\overline{\Psi}=\Psi^{\dag} \gamma^0.$ First, let us make few transformations.

The volume element in the above action is defined with respect to a moving frame associated to the surface
\begin{equation}\label{3dMetric}
\vec{X}(q_1, q_2, q_3)=\vec{x}(q_1, q_2)+q_3 \vec{N}(q_1, q_2), 
\end{equation}
where $\vec{N}(q_1, q_2)$ is a vector normal to the surface. This coordinate system facilitates the probing of the three dimensional space $\vec{X}(q_1, q_2, q_3)$ from the viewpoint of a two dimensional surface $\vec{x}(q_1, q_2).$ For the volume element we obtain \cite{daCosta}
\begin{equation}
dV=f dS dq_3,  
\end{equation}
where
\begin{equation}\label{eq:f}
f=1-2M q_3 + K q_3^2
\end{equation}
and $dS=\sqrt{g}dq_1 dq_2.$ Hereafter $g_{ij}$'s are the metric components and $g$ is the determinant of the metric associated with the surface $\vec{x}(q_1,q_2);$ $M$ is the Mean and $K$ the Gaussian curvature of this surface,  respectively.

In this construction the spinor wavefunction is normalized according to
\begin{equation}
\int f dS dq_3 \overline{\Psi} \gamma^0 \Psi =1
\end{equation}
and in order to facilitate the separation of the dynamics into two parts, surface and orthogonal to it, we rewrite the action with the substitution
\begin{equation}
\Psi = \frac{\psi}{\sqrt{f}}.
\end{equation}
The normalization condition for the new spinor wavefunction $\psi$ is the following
\begin{equation}
\int dS dq_3 \overline{\psi} \gamma^0 \psi =1
\end{equation}
and the action is
\begin{eqnarray}
\nonumber S&=&\int \sqrt{f} dS dq_3 \; \overline{\psi} \left[ -i\hbar v_F \sum \gamma^a D_a  + m(q_3)v_F^2 + V_{\perp}(q_3) \right] \frac{\psi}{\sqrt{f}}.
\end{eqnarray}
Now we pull the factor $f^{-1/2}$ through the gauge covariant derivative. For the corresponding equation of motion we obtain
\begin{eqnarray}
\nonumber  && \left\{ -i\hbar v_F \gamma^3 \partial_3 + m(q_3)v_F^2 + V_{\perp}(q_3) \right\} \psi \\ 
\nonumber && - i\hbar v_F \gamma^3 \mathcal{D}_3 \psi  -i\hbar v_F \gamma^3 \left[ f^{1/2} \partial_3 (f^{-1/2}) \right] \psi  \\
&& -i\hbar v_F  \sum_{a=0}^2 \gamma^a  f^{1/2} D_a \frac{\psi}{\sqrt{f}} =0,
\end{eqnarray}
where $ \mathcal{D}_3$ is the remaining part of the gauge covariant derivative (\ref{eq:D}) after the partial derivative $\partial_3$ has been removed. Here using the definition of $f$ (\ref{eq:f}), 
\begin{equation}
 f^{1/2} \partial_3 (f^{-1/2}) =  f^{1/2} M.
\end{equation}

Now the tetrad fields, in the definition of the curved gamma matrices, constructed from the metric tensor  (\ref{3dMetric}) $w_a=e_a^\mu d q_\mu$ can be read off

\begin{eqnarray}
w_0&=&dt ,\\
w_1&=&\sqrt{G_{11}}dq_1 ,\\
w_2&=&\sqrt{G_{22}}dq_2 ,\\
w_3&=&dq_3 , 
\end{eqnarray}
where $G_{ij}$ are the corresponding metric components of (\ref{3dMetric}).
Note $e^3_\mu $ is constant which means that $\gamma^3$ is constant.

We are now ready to take into account the effect of the constraining potential $V_{\perp}(q_3) $ since when it applies the wavefunction is compressed between two steep potential barriers on both sides of the surface. The value of the wavefunction will be significantly different from zero only for very small range of values around $q_3=0.$ Therefore, we take the limit $q_3 \rightarrow 0$ 
in all coefficients in the above equation except the terms containing $ m(q_3)$ and $V_{\perp}(q_3) .$ In this limit  the tetrad constructed from the full metric tensor $e^a_\mu (q_1,q_2,q_3) \to e^a_\mu (q_1,q_2)$ and the curved gamma matrices are defined with respect to the surface metric $G_{ij}\to g_{ij}$
\begin{equation}\label{gamma_curved}
\gamma^a(q_1, q_2)=e^a_\mu (q_1, q_2) \gamma^\mu,  
\end{equation}
where the tetrad field  $w_a=e_a^\mu d q_\mu$ is given with respect to the surface metric $g_{ij}$

\begin{eqnarray}
w_0&=&dt ,\\
w_1&=&\sqrt{g_{11}}dq_1 ,\\
w_2&=&\sqrt{g_{22}}dq_2 ,\\
w_3&=&dq_3 .
\end{eqnarray}

In the limit we also have $f \rightarrow 1$ and $\mathcal{D}_3 \rightarrow - i \frac{e}{c}A_3$.
As a result we obtain the following equation
\begin{eqnarray}
\nonumber  && \left\{ -i\hbar v_F \gamma^3 \partial_3 + m(q_3)v_F^2 + V_{\perp}(q_3) \right\} \psi  -  \gamma^3 \frac{e \hbar v_F}{c}A_3 \psi  \\
&&  -i\hbar v_F  \sum_{a=0}^2 \gamma^a  D_a \psi -i\hbar v_F \gamma^3 M \psi =0.
\end{eqnarray}
In Coulomb gauge $A_3=0.$ The wavefunction is separable $\psi=\chi_N \chi_T$ into normal to the surface $\chi_N$ and tangential $\chi_T$ part. Since $e^3_\mu $ is constant, so is $\gamma^3$ and the equation is separable
\begin{eqnarray}\label{dirac_2d_curved}
i\hbar v_F  \gamma^0  \partial_t \chi_T =  -i\hbar v_F  \sum_{a=1}^2 \gamma^a  D_a \chi_T -i\hbar v_F \gamma^3 M \chi_T ,     \\
i\hbar v_F  \gamma^0  \partial_t \chi_N =   -i\hbar v_F \gamma^3 \partial_3 + m(q_3)v_F^2 + V_{\perp}(q_3)  \chi_N . 
\end{eqnarray}
Here $D_a$ are gauge covariant derivatives defined with respect to the two dimensional metric of the surface $\vec{x}(q_1,q_2).$ The second equation in the case of graphene should be solved in the nonrelativistic limit since it describes the off surface dynamics. 

It turns out that the nonrelativistic system tends to extract more information from the manifold than the relativistic one since the relativistic fermion system only probes the surrounding space to first order in the derivatives of the wavefunction, whereas the Schr\"odinger fermions or bosons probe it to second order.  More specifically, the Gauss curvature $K$ appears only non-relativistically and is related through the Gauss-Bonet theorem to the topology class of the manifold. In the nonrelativistic limit this information is returned as higher derivatives of the wavefunction. As a result the Gauss curvature does not appear explicitly in the confining potential for the relativistic case.

\section{Graphene with a bump} 

Now, let us explore a particular example, namely a bump on the surface of graphene. The bump has a Gaussian shape and is rotationally invariant. The reason we explore this geometry is the fact that it has been discussed previously\cite{vozmediano} and we can compare our results which depend on the confining procedure with the results obtained by postulating the two dimensional Dirac equation to which the carriers of graphene are subjected. It turns out that due to the confining procedure an extra term appears which is proportional to the Mean curvature of the surface.

 \begin{figure}[ht]
\begin{center}
\includegraphics[scale=0.4]{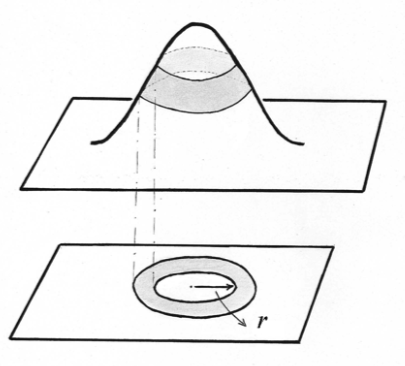}
\caption{\label{fig:bump} A bump on the graphene surface affects local density of states (LDOS) as inferred by (\ref{eq:H_bump}). The shaded region is approximately where the Mean curvature is at maximum and the LDOS should be affected the most.}
\end{center}
\end{figure}

In polar coordinates the surface is given by
\begin{equation}
z(r)=A e^{-r^2/b^2}
\end{equation}
and is asymptotically flat, $\lim_{r \to \infty} z(r)=0$ (see Fig. \ref{fig:bump}).  We will use this in order to check if the equation we obtain has the correct limit.
Let us denote
\begin{equation}
dz^2=\left( \frac{dz}{dr}\right)^2 dr^2=\alpha f(r) dr^2 , 
\end{equation}
which leads to the following for the line element
\begin{equation}
ds^2=dr^2+r^2d\theta^2+dz^2=\left[1+\alpha f(r) \right] dr^2 + r^2 d \theta,
\end{equation}
where $\theta \in [0, 2 \pi ] $ is the polar angle. The metric is
\begin{equation}\label{g_bump}
g_{\mu \nu}=\left(\begin{array}{cc}1+\alpha f(r) & 0 \\0 & r^2\end{array}\right).
\end{equation} 

Before we derive the Dirac equation for the bump we will focus on the flat case which will provide the asymptotic. The flat case is obtained by taking $\alpha=0.$ Then the metric reduces to
\begin{equation}\label{g_flat}
g_{\mu \nu}=\left(\begin{array}{cc}1 & 0 \\0 & r^2\end{array}\right),
\end{equation} 
which yields one possible choice for the moving frame $g_{\mu \nu}=e^a_\mu e^b_\nu \delta_{ab}:$
\begin{equation}
e^a_{\mu}=\left(\begin{array}{cc}1 & 0 \\0 & r\end{array}\right).
\end{equation} 
The affine connection $\Gamma^\alpha_{\beta \gamma}$ for this metric has the nonvanishing elements
\begin{equation}
\Gamma^r_{r r}=0, \quad \Gamma^r_{\theta \theta}=-r, \quad \Gamma^\theta_{r \theta}=\frac{1}{r} , 
\end{equation}
which play a role in computing the spin connection \cite{weinberg} $\Omega_\mu$ 
\begin{equation}
\Omega_\mu = \frac14 \gamma_a \gamma_b e^a_\lambda g^{\lambda \sigma} \left( \partial_\mu e^b_{\sigma} - \Gamma^{\lambda}_{\mu \sigma}e^b_\lambda \right).
\end{equation}
The spin connection takes part in the two dimensional Dirac equation on the metric $g_{\mu \nu}$ as follows
\begin{equation}
i \vec{\gamma}^\mu (\partial_\mu + \Omega_\mu) \psi=0,
\end{equation}
where $\vec{\gamma}=(\gamma^0, v_F \gamma^i)$;  see the Appendix. 

Finally the flat hamiltonian is 
\begin{eqnarray}
 H_{flat}&=&-i\sigma_3\partial_t + \left(\begin{array}{cc}0 & \partial_r + i \frac{\partial_\theta}{r}+\frac{1}{2r} \\\partial_r - i \frac{\partial_\theta}{r}+\frac{1}{2r} & 0\end{array}\right) . 
\end{eqnarray} 

Now that we know the asymptotic limit we may derive the corresponding hamiltonian for the Gaussian bump keeping in mind that due to confinement we have an extra term in the hamiltonian $-i\hbar v_F \gamma^3 M$ as indicated by equation (\ref{dirac_2d_curved}). This extra term in the 2 representation we are working reduces to
\begin{equation}\label{eq:geompotM}
V = -i\hbar v_F  M 
\end{equation} 
due to the properties of the $\gamma$ matrices \cite{Jensen, pnueli, leinaas}. In the curved geometry
\begin{eqnarray}\label{M}
M&=&\frac12 (\kappa_1+\kappa_2)
= -\frac12 \left[ \frac{\sqrt{\alpha f} }{r \sqrt{1+\alpha f}} + \frac{\alpha \frac{df}{dr} }{2\sqrt{\alpha f} (1+\alpha f)^{3/2} }   \right],
\end{eqnarray}
where $\kappa_i$ are the principal curvatures of the surface.

For the curved metric (\ref{g_bump}) the moving frame can be given by
\begin{equation}
e^a_{\mu}=\left(\begin{array}{cc}\sqrt{1+\alpha f } & 0 \\0 & r\end{array}\right) , 
\end{equation} 
which together with the affine connection
\begin{equation}
\Gamma^r_{r r}=\frac{\alpha \frac{df}{dr}}{2(1+\alpha f)}, \quad \Gamma^r_{\theta \theta}=-\frac{r}{1+\alpha f}, \quad \Gamma^\theta_{r \theta}=\frac{1}{r} , 
\end{equation}
yields for the spin connection
\begin{eqnarray}
\Omega_r&=&0, \\
\Omega_{\theta}&=&\frac12 \gamma_2 \gamma_1 \frac{1}{\sqrt{1+\alpha f} }=-i\sigma_3 \frac12 \frac{1}{\sqrt{1+\alpha f} }.
\end{eqnarray}
Finally for the Dirac equation describing the carriers in the curved graphene sheet we have
\begin{equation}
\left[ i \vec{\gamma}^\mu (\partial_\mu + \Omega_\mu) -i\hbar v_F  M\right] \psi=0,
\end{equation}
where $\vec{\gamma}=(\gamma^0, v_F \gamma^1, v_F \gamma^2)$ is defined through the moving frame (\ref{gamma_curved}). The corresponding hamiltonian is
\begin{eqnarray}\label{eq:H_bump}
 H_{bump}&&=-i\sigma_3\partial_t -i\hbar v_F  M\\
\nonumber && + \left(\begin{array}{cc} 0 & \frac{\partial_r}{\sqrt{1+\alpha f} } + i \frac{\partial_\theta}{r}+A_{\theta} \\ \frac{\partial_r}{\sqrt{1+\alpha f} } - i \frac{\partial_\theta}{r}+A_{\theta} & 0\end{array}\right),
\end{eqnarray} 
where the magnetic field associated with the gauge field
\begin{equation}
A_{\theta}=\frac{\Omega_{\theta}}{r}=\frac{1}{2r\sqrt{1+\alpha f}}
\end{equation}
is substantial
\begin{equation}
B_z=-\frac{1}{r} \partial_r (r A_{\theta})
\end{equation}
and for the typical size of the bumps on graphene varies between 0.5 - 3 T.
It is straightforward to check the asymptotic form $\alpha=0$ of this hamiltonian and to see that indeed
\begin{equation}
H_{bump} \to H_{flat}
\end{equation}
when the bump flattens.

One of the most obvious consequences of this geometric potential (\ref{eq:geompotM}) is that it shifts the Fermi energy. Thus controlling graphene's electronic properties can be achieved not only by doping and creating imperfections in the lattice, but also by {\it bending} \cite{avadh&me}.

\section{Graphene in the context of a wormhole geometry}

Since we have developed an effective treatment of the massless carriers in graphene for the low-energy electronic properties of curved sheets in the continuum limit, we proceed to apply it to the wormhole geometry\cite{thorn}. Our motivation stems from one of the biggest gaps in modern physics, namely the relation between Quantum Mechanics and General Theory of Relativity. Constrained quantum systems present us with the opportunity to explore the quantum mechanical consequences of curvature.

The metric of the wormhole is given by

\begin{equation}
ds^2=-c^2dt^2 + dl^2+(b^2+l^2)(d\phi^2+\sin^2 \phi d\theta^2),
\end{equation}
where $t$ is the proper time of a static observer, $l=\pm \sqrt{r^2+b_0^2}$ is the proper radial distance at constant time, $b=b(l)$ is the shape function of the wormhole [$b(0)=b_0$ is the radius of the throat of the wormhole] and $(\theta, \phi)$ are spherical polar coordinates.

\begin{figure}[ht]
\begin{center}
\includegraphics[scale=0.4]{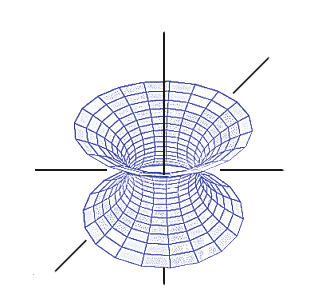}
\caption{\label{fig:catenoid} Any two-dimensional section (catenoid) of a three dimensional wormhole geometry. A large graphene sheet folded like this may serve as a solid state experimental realization of transport of relativistic particles trough a wormhole. In the process of folding the hexagonal symmetry of the underlying lattice should not be disrupted.}
\end{center}
\end{figure}

As is usually the case with quantum systems we are interested in eigenmodes and we fix the time $t={\rm const.}$ We also consider the case $\phi=\pi/2$ which represents an equatorial cross-section of a three-dimensional wormhole at constant time. The line element therefore becomes
\begin{equation}
ds^2=dl^2+(b^2+l^2)d\theta^2,
\end{equation}
which is precisely equivalent to the line element of the catenoid\cite{rossen}
\begin{equation}
ds^2=\frac{r^2}{r^2+b_0^2}dr^2+r^2 d\theta^2.
\end{equation}
Note that at any arbitrary cross-section of the three-dimensional wormhole say $\phi=\phi_0,$ the line element is
\begin{equation}
ds^2=\frac{r^2}{r^2+b_0^2}dr^2+\epsilon^2 r^2 d\theta^2,
\end{equation}
where $\epsilon^2=\sin^2 \phi \in [0,1].$ For the catenoid this will mean only rescaling the radius from $r$ to $\epsilon r.$ The catenoid with the biggest radius corresponds to the equatorial cross-section, see Fig. \ref{fig:catenoid}.

In cylindrical coordinates $(z,r, \theta)$ a two dimensional cross-section of a wormhole is given by
\begin{equation}
z(r)=\pm \ln\left[\frac{r}{b_0}+\sqrt{\frac{r^2}{b_0^2}-1}  \right].
\end{equation}
The principal curvatures $\kappa_1$ and $\kappa_2$ of this surface are
\begin{equation}
\kappa_1=\frac{1}{b_0} {\rm sech}^2 \frac{z}{b_0}, \qquad \kappa_2=-\kappa_1.
\end{equation}

This means that any two dimensional cross-section of a wormhole is a minimal 
surface
\begin{equation}
M=\frac12 (\kappa_1+\kappa_2)=0.
\end{equation}

Since the Fermi velocity in graphene is 0.3\% of the velocity of light, the mixing of space and time coordinates is negligible. Therefore, the two dimensional catenoid section of the wormhole geometry can be realized through folded graphene providing us with a genuine solid state experiment.

The constrained fermion dynamics we derived for the purposes of obtaining the electronic properties of curved graphene can now be combined with the geometry of the catenoid as an experimental solid state proposition testing the transmission of material particles through a wormhole. Indeed graphene can be shaped as catenoid provided the sheet is big enough so that while bending it one does not destroy the hexagonal symmetry of the underlying lattice and lose the relativistic description\cite{guinea}. The destruction of the hexagonal symmetry can be inferred by the splitting of the G peak in the Raman spectra of graphene\cite{Ferrari, avadh}.

It appears that due to the vanishing of the mean curvature, no extra potential subjects the relativistic particles heading for a transition ``on the other side" of the wormhole. The governing equation is the free Dirac equation in curvilinear coordinates.

\section{Conclusions}

We have demonstrated that due to the essentially three dimensional nature of the carriers of graphene, using a purely two dimensional relativistic description to derive their properties contradicts the Heisenberg principle. In order to obtain the correct relativistic equation we employed a gradual confining procedure which produced a geometric potential in the equation of motion. As a result, the possibility to open a band gap in a purely mechanical way, simply by {\it bending} the material, is demonstrated. The result presented here also connects to the properties of quantum systems in the curved space-time of general relativity. It is interesting to note that due to the coupling of mechanical and electronic degrees of freedom through the geometric gauge potential (which for nanoscale radii of curvature is substantial)
\[
V=- i\hbar v_{F} M ,
\]
mechanical vibrations of graphene can induce resonance electronic transitions, since the band gap is proportional to the curvature created by the mechanical oscillation itself. This can result in a new class of MEMS and NEMS  based on the geometric properties of graphene.

Even more intriguing is the observation that when a graphene membrane in MEMS and NEMS is set to oscillate this would result in a time-dependent Mean curvature $M \sim \sin{(2 \pi f_{res} t)}$ term, which can mechanically drive transitions between and within electronic bands\cite{bachtold}. Strain induced modification of the electronic properties or ``straintronics" is clearly an important field of study in graphene.  Moreover, the band gap of the material can be tuned in the terahertz region where powerful sources of coherent radiation are much needed in  industrial applications involving security aspects and medical scans.

\bigskip

\section*{Appendix}

In 2+1 flat dimensions the fundamental 2 representation of the $\gamma$ matrices is satisfied by
\begin{equation*}
\gamma_0=-i\sigma_3, \quad \gamma_i=-\sigma_i
\end{equation*}
for $i=1,2.$
Here $$\sigma_1=\left(\begin{array}{cc}0 & 1 \\1 & 0\end{array}\right),\qquad \sigma_2=\left(\begin{array}{cc}0 & -i \\i & 0\end{array}\right), \qquad \sigma_3=\left(\begin{array}{cc}1 & 0 \\0 & -1\end{array}\right). $$

They obey the following relations
\begin{eqnarray*}
&& \{ \gamma^a , \gamma^b  \}=\eta^{ab},\\
&&\gamma^a \gamma^b=\eta^{ab}+\epsilon^{abc} \gamma^c,\\
&&{\rm tr} (\gamma^a \gamma^b \gamma^c)=-2 \epsilon^{abc}, \\
&&\gamma_5= \gamma_0 \gamma_1 \gamma_2,\\
&&[ \gamma_5, \gamma_{a} ] = 0
\end{eqnarray*}

\section*{Acknowledgments} This work was supported in part by the U.S. Department of Energy. V.A. also acknowledges support by ICTP, Trieste, Italy where this paper was completed.

\end{document}